\documentclass[%
 reprint,
 amsmath,amssymb,
 aps,
floatfix,
]{revtex4-2}
\newcommand{\comment}[1]{}

\newcommand{\figonecols}[1]{\includegraphics[width=\linewidth]{#1}	}

\usepackage[utf8]{inputenc}
\usepackage[section]{placeins}
\usepackage{float}
\usepackage{graphicx}
\usepackage{multirow}
\usepackage{xcolor}
\usepackage{color}
\usepackage{braket}
\usepackage{ulem}

\usepackage[margin=0.5in]{geometry}
\usepackage{amsmath}
\usepackage[colorlinks=true,linkcolor=black,anchorcolor=black,citecolor=black,filecolor=black,menucolor=black,runcolor=black,urlcolor=black]{hyperref}
\begin{document}
\title{Long Lived Electronic Coherences in Molecular Wave Packets Probed with Pulse Shape Spectroscopy}

\author{Brian Kaufman,$^{1}$ Philipp Marquetand,$^{2}$ Tam\'as Rozgonyi,$^{3}$ and Thomas Weinacht$^{1}$}

\affiliation{$^{1}$Department of Physics and Astronomy, Stony Brook University, Stony Brook, New York 11794-3800, USA}

\affiliation{$^{2}$University of Vienna, Faculty of Chemistry, Institute of Theoretical Chemistry, W\"ahringer Strasse 17, 1090 Wien, Austria}

\affiliation{$^{3}$Wigner Research Centre for Physics, P.O. Box 49, H-1525 Budapest, Hungary}

\date{\today}

\begin{abstract}
We explore long lived electronic coherences in molecules using shaped ultrafast laser pulses to launch and probe entangled nuclear-electronic wave packets. We find that under certain conditions, the electronic phase remains well defined despite vibrational motion along many degrees of freedom. The experiments are interpreted with the help of electronic structure calculations which corroborate our interpretation of the measurements.  
\par
\end{abstract}
\maketitle
\maketitle
\section{Introduction}
The motion of electrons in photoexcited molecules drives many basic light-driven processes in physics, chemistry and biology.  From solar cells to photodissociation and photosynthesis, electronic dynamics play a fundamental role in molecular transformation, and can determine what the final products are \cite{worner2017charge}. Electronic dynamics can be described in terms of wave packets - coherent superpositions of electronic eigenstates, whose evolution is dictated by the relative phase between states \cite{PhysRevLett.126.133002,haessler2010attosecond,okino2015direct}.  While this phase relationship (``electronic coherence") remains well defined in atoms for many cycles \cite{weinacht1998measurement, PhysRevLett.64.2007, PhysRevLett.92.133004},  in molecules it is complicated by the motion of the nuclei, which are entangled with the electrons - i.e. the full wave function generally cannot be written as a product of electronic and nuclear wave functions, and the entanglement of the wave function typically leads to a rapid decay in the electronic coherence if one does not perform nuclear coordinate resolved measurements.  

The loss of electronic coherence as a consequence of electron-nuclear coupling can be seen by considering the total wave function as a Born-Oppenheimer or Born-Huang expansion \cite{born1985international}:
\begin{equation}
    \Psi(\boldsymbol{r},\boldsymbol{R},t)=\sum_n a_n \chi_n(\boldsymbol{R},t) \psi_n(\boldsymbol{r};\boldsymbol{R})
    \label{eq:wpkt}
\end{equation}
where $\boldsymbol{r}$ and $\boldsymbol{R}$ represent the electronic and nuclear degrees of freedom respectively, $\psi_n(\boldsymbol{r};\boldsymbol{R})$  represents the n$^{th}$ electronic eigenstate of the molecule, $a_n$ is the complex amplitude of the n$^{th}$ state, and $\chi_n(\boldsymbol{R},t)$ represents the (normalized) time-dependent nuclear wave function in the n$^{th}$ electronic state \footnote{We note that this expansion is slightly different from the standard Born-Huang expansion, for which the $c_n$ coefficient is included in the vibrational wave function $\chi_n(\boldsymbol{R},t)$. We have separated c$_n$ out here since it allows us to use normalized vibrational wave functions and separate the parts of the wave function that are laser dependent from those that are not.}. 
As shown in earlier work \cite{Vibok}, the excitation of such a wave packet leads to an electronic coherence which is given by the off diagonal element of the density matrix, and can be written as:
\begin{equation}
    \rho_{12}(\boldsymbol{R},t)=a_1 \chi_1(\boldsymbol{R},t) a_2^* \chi_2^*(\boldsymbol{R},t) e^{i(\omega_2(\boldsymbol{R})-\omega_1(\boldsymbol{R}))t}
    \label{eq:coherence}
\end{equation}
where $\hbar \omega_1(\boldsymbol{R}) =V_1(\boldsymbol{R})$ and $\hbar \omega_2(\boldsymbol{R}) =V_2(\boldsymbol{R})$, with V$_1(\boldsymbol{R})$ and V$_2(\boldsymbol{R})$ being the potential energies of states 1 and 2 as a function of nuclear coordinate respectively.
Calculations and measurements over the past two decades have established rather short lifetimes for $\rho_{12}(t)$ - less than 10 fs - \cite{arnold2017electronic, Vibok,franco2008femtosecond,scheidegger2022search,hwang2004electronic,kamisaka2006ultrafast,PhysRevLett.118.083001} due to loss of vibrational wave function overlap (the product of $\chi_1(\boldsymbol{R},t)$ and $\chi_2(\boldsymbol{R},t)$ in Eq. \ref{eq:coherence}), different rates of phase advance on states 1 and 2 (dephasing of $e^{i(\omega_2(\boldsymbol{R}) - \omega_1(\boldsymbol{R}))t}$ in Eq.~\ref{eq:coherence}), and internal conversion (decay of $a_1$ and $a_2$ in Eq.~\ref{eq:coherence} - i.e. $a_n \rightarrow a_n(t)$) \cite{Vibok,arnold2017electronic}. 
This has led to a significant debate over the role that electronic coherences play in photosynthesis and other natural processes driven by light absorption \cite{duan2017nature,maiuri2018coherent}.
Here we build on recent work \cite{kaufman2023PRL} and explore electronic coherences in molecules when the electronic states in question are approximately parallel and the normally entangled wave function can be roughly factored.  This mitigates vibrational dephasing and maintains vibrational overlap for longer times, allowing for $\rho_{12}(\boldsymbol{R},t)$ to survive even in the face of averaging over $\boldsymbol{R}$.

\section{Measurement Approach}

\begin{figure}[htbp]
    \centering	
    \figonecols{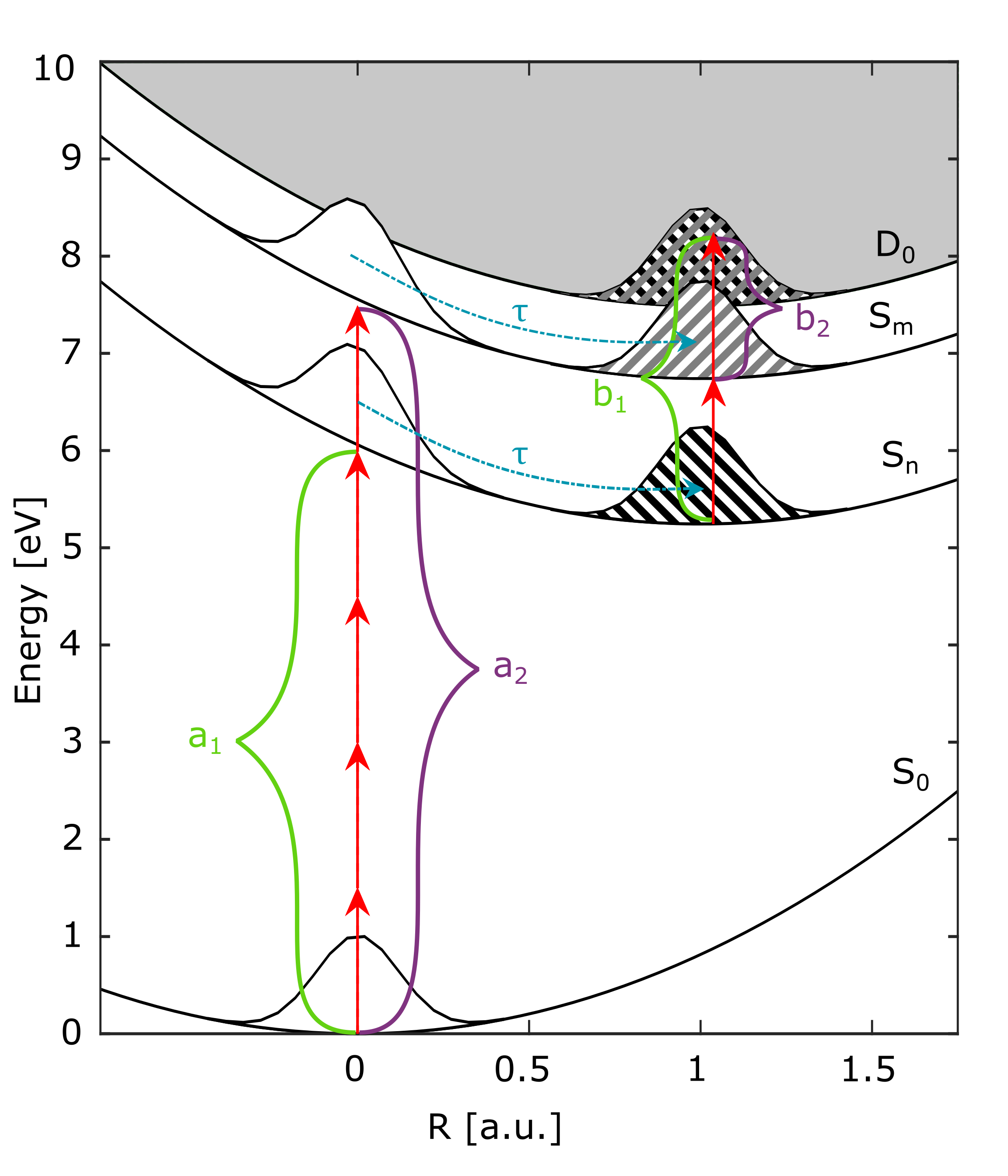}
    \caption{Cartoon potential energy curves for a four state system with ground state $0$, two excited states $1$ and $2$ and ionic state $3$. The red arrows describe the multiphoton coupling photon orders and the coefficients $a$ and $b$ describe the multiphoton coupling strengths, where $a$ describes the excitation and $b$ the ionization.}
    \label{fig:cartoon}
\end{figure}

In our experiments, a pump pulse which can be written as $E_{\text{pu}}(t)=E_0(t)\cos(\omega_0 t)$, prepares a wave packet described by Eq.~\ref{eq:wpkt} via multiphoton absorption. This is illustrated in Fig.~\ref{fig:cartoon}, which shows N photon absorption to state 1 and N+1 photon absorption to state 2, with N=4 for the specific experiments described here. 

Impulsive multiphoton ionization with a phaselocked probe pulse at time $\tau$ ($E_{\text{pr}}(t)=E'_0(t-\tau)\cos(\omega_0 (t-\tau)-\phi)$) produces an ionization yield that can be written as:
\begin{equation} Y(t,\tau,\phi)=|a_1|^2|b_1|^2+|a_2|^2|b_2|^2\\ +b_1 b_2^*  \int d\boldsymbol{R} \rho_{12}(\boldsymbol{R},t) + c.c.
    \label{eq:dication_yield}
\end{equation}
where $b_1$ and $b_2$ represent $m$ and $m-1$ photon (with $m$=2 in this work) ionization amplitudes, which are proportional to the $m^{th}$ and $(m-1)^{th}$ powers of the probe pulse field respectively:
\begin{subequations}
\begin{align}
    b_1(t) &= Q_{1\text{f}}\big(E_\text{pr}(t)\big)^m \\
    b_2(t) &= Q_{2\text{f}}\big(E_\text{pr}(t)\big)^{(m-1)}
    \label{Aeq:bn}
\end{align}    
\end{subequations}

Here $Q_{1\text{f}}$ and $Q_{2\text{f}}$ represent field independent multi-photon matrix elements that can be written in terms of sums over off resonant intermediate states \cite{lunden2014model,kaufman2022numerical}.  \comment{This leads to a measured yield as a function of pump-probe delay and phase ($Y_{pp}(\tau,\phi)$) which can be written as (see appendix for details):
\begin{align}
      Y_{pp}(\tau,\phi)&=|a_1|^2|Q_{1f}(E'_0)^m|^2+|a_2|^2|Q_{2f}(E'_0)^{m-1}|^2 \label{eq:dication_yield2} 
      \\ 
      &+ Q_{1\text{f}} Q^*_{2\text{f}} (E'_0)^{(2m-1)}e^{i\phi}  \int d\boldsymbol{R} \rho_{12}(\boldsymbol{R},\tau) + c.c. \nonumber
  \end{align}}
If one writes a delay dependent phase on the probe pulse given by $\phi=\phi_L-\omega_L \tau$, where $\omega_L$ is a locking frequency discussed below, then the yield as a function of pump probe delay and phase can finally be written as (see appendix for details):
\begin{align}
\label{eq:final_yield_phase_tau}
    Y(\tau,\phi) &= |a_1|^2|Q_{1f}(E'_0)^m|^2+|a_2|^2|Q_{2f}(E'_0)^{m-1}|^2 \\
    &+a_1a_2^*Q_{1\text{f}}Q^*_{2\text{f}}(E'_0)^{(2m-1)} \int d\boldsymbol{R} \chi_1(\boldsymbol{R},\tau)\chi_2^*(\boldsymbol{R},\tau) \nonumber
    \\
    &e^{i((\omega_{21}(\boldsymbol{R})-\omega_L)\tau+\phi_L)} + c.c.\nonumber
\end{align}
\section{Experimental Apparatus}
Eq.~\ref{eq:final_yield_phase_tau} highlights the fact that excitation and ionization with a phaselocked pulse pair where $\phi$ can be controlled allows for measurements of the coherence between states 1 and 2, provided that integration over $\boldsymbol{R}$ does not wash it out. The generation of such a controllable phaselocked pulse pair can be achieved through the use of an optical pulse shaper. 
  
Our experiments make use of an amplified titanium sapphire laser system, which produces transform limited $30$~fs laser pulses with an energy of $1$~mJ at a repetition rate of $1$~kHz.  The pulses are spectrally broadened in a $2.1$~m $450$~$\mu$m-core stretched hollow core fiber filled with $600$~Torr of ultra-high purity Argon gas. The input spectrum centered around $780$~nm is blue-shifted to $750$~nm and broadened from $600-900$~nm \cite{nisoli_1996_GenerationHighEnergy,hagemann_2013_SupercontinuumPulseShaping,catanese_2021_AcoustoopticModulatorPulseShaper}. 

An acousto-optic modulator (AOM) based pulse shaper \cite{dugan1997high} is used for compression, characterization, and shaping of these ultrabroadband pulses. In the pulse shaper the AOM is used as a spectral mask, $M(\omega)$, to shape the pulse in the frequency domain by placing it in the Fourier plane of a zero dispersion stretcher \cite{weiner_2000_FemtosecondPulseShaping}. The shaped electric field, $E'(\omega)$, is a product of the acoustic mask, $M(\omega)$, and the unshaped field, $E(\omega)$: $E'(\omega) = M(\omega)E(\omega)$. Using a phase mask modelled by a Taylor-series expansion up to fourth order dispersion combined with the residual reconstructed phase from a pulse-shaper-assisted dispersion scan (PS-DSCAN) \cite{catanese_2021_AcoustoopticModulatorPulseShaper}, the broadened spectrum is compressed to near transform limit, $7$~fs. The pulse can be further characterized temporally using a pulse-shaper-assisted, second harmonic generation collinear frequency resolved optical gating technique or PS-CFROG \cite{trebino_1997_MeasuringUltrashortLaser,amat-roldan_2004_UltrashortPulseCharacterisation} which confirms the $7$~fs pulses.

For the experiments we explore multiple mask functions. One is modelled by a Gaussian to narrow the optical spectrum and perform measurements for different optical frequencies. This equation is written as:
\begin{equation}
    M(\omega) = A\exp\bigg(\frac{(\omega-\omega_c)^2}{\Delta_{\omega}^2}\bigg) = M_{WS}
    \label{eq:maskWS}
\end{equation}
where $A$ is the overall amplitude, $\omega_c$ is the optical frequency at which the chosen window is centered, and $\Delta_{\omega}$ is the width of the narrowed optical bandwidth. Another mask function generates a pulse pair with independent control over both the pump-probe delay ($\tau$) and the relative phase between the pulses ($\phi_L$) while establishing a locking frequency ($\omega_L$) and is written as:
\begin{equation}
    M(\omega) = A_T\Big(1+A_R e^{i(\omega-\omega_L)\tau+i\phi_L}\Big) = M_{PP} \\
    \label{eq:maskPP}
\end{equation}
where $A_T$ is the overall amplitude and $A_R$ is the relative pump-probe amplitude such that $E'_0(t)=A_RE_0(t)$.  The relative phase between pulses described by the probe pulse, $\phi$, can be expressed in terms of $\phi_L$ and $\omega_L$ by $\phi=\phi_L-\omega_L \tau$. This leads to the two different phase sensitive measurements that we carried out, which were to directly scan $\phi_L$ for a fixed value of $\tau$ (phase measurement), and to vary $\tau$ (pump-probe measurement). 

Finally, we note that combining the two mask functions, $M(\omega)=M_{WS}M_{PP}=M_{WPP}$, allows us to perform a pump-probe measurement for narrowed spectra at different central frequencies.

These shaped pulses are then focused in an effusive molecular beam inside a vacuum chamber with a base pressure of $\sim 10^{-10}$~Torr, raising the working pressure to about $\sim10^{-6}$~Torr.  The molecules are ionized by the laser pulses, with peak intensities of up to $\sim10^{13}$~W/cm$^2$). The electrons generated by ionization are velocity map imaged to a dual-stack microchannel plate (MCP) and phosphor screen detector using an electrostatic lens.  The light emitted by the phosphor screen at each position is recorded by a CMOS camera.  The camera measurements are inverse Abel transformed to reconstruct the three-dimensional momentum distribution of the outgoing electrons and the photoelectron spectrum (PES). The following measurements describe the photoelectron spectrum or yield as a function of the various pulse shaper parameters.

\comment{Our measurements make use of an amplified Ti:sapphire laser system generating $1$~mJ transform limited pulses of $30$~fs duration, centered at a wavelength of $780$~nm, and operating at a $1$~kHz repetition rate. The pulses are spectrally broadened using self-phase modulation in a $2.1$~m stretched-hollow core fiber (S-HCF) filled with 600 Torr of ultra-high purity Argon gas. The S-HCF produces a slightly blueshifted spectrum (central wavelength of $750$~nm), extending from $600-900$~nm \cite{nisoli_1996_GenerationHighEnergy,hagemann_2013_SupercontinuumPulseShaping,catanese_2021_AcoustoopticModulatorPulseShaper}. The broadened spectrum is compressed to near transform limit, $7$~fs, using a phase mask modelled by a Taylor-series expansion up to fourth order dispersion combined with the residual reconstructed phase from a pulse-shaper-assisted dispersion scan (PS-DSCAN) \cite{catanese_2021_AcoustoopticModulatorPulseShaper}. This is then characterized using pulse-shaper-assisted, second harmonic generation collinear frequency resolved optical gating (PS-CFROG) \cite{trebino_1997_MeasuringUltrashortLaser,amat-roldan_2004_UltrashortPulseCharacterisation}.
The shaped pulses are focused in an effusive molecular beam inside a vacuum chamber with a base pressure of $\sim 10^{-10}$ Torr, raising the working pressure to about $\sim 10^{-6}$ Torr.  The molecules are ionized by the laser pulses, with peak intensities of up to $\sim 10^{13}$~W/cm$^2$). The electrons generated by ionization are velocity map imaged to a dual-stack microchannel plate (MCP) and phosphor screen detector using an electrostatic lens.  The light emitted by the phosphor screen at each position is recorded by a CMOS camera.  The camera measurements are inverse Abel transformed to reconstruct the three-dimensional momentum distribution of the outgoing electrons and the photoelectron spectrum (PES).}

\section{Electronic Structure Calculations}

The equilibrium geometry of the ground electronic state and singlet excited state energies of thiophene at this geometry were determined by electronic structure calculations in Ref.~\cite{kaufman2022numerical}. The geometry optimization was performed by density funtional theory using the Gaussian program package~\cite{G09} with the B3LYP functional~\cite{b3lyp1,b3lyp2} and aug-cc-pVTZ basis set~\cite{dftbasis} for all the atoms. The state energies were determined at the multistate complete active space 2nd order perturbation theory (MS-CASPT2) level~\cite{MSCASPT2} for 30 states with the help of the Open-Molcas 20.10 program package~\cite{Molcas2010}. The active space consisted of 10 electrons and 11 orbitals. (The shape of the active orbitals are displayed in Fig. 10 of Ref.~\cite{kaufman2022numerical}) For further details of the computations see Ref.~\cite{kaufman2022numerical}.


\section{Measurements}
In order to create and probe electronic coherences as outlined in the introduction, we first need to drive multiphoton excitation and ionization (Resonance Enhanced MultiPhoton Ionization, or REMPI) and then drive two resonances simultaneously in order to see interference.  

\subsection{Multiphoton Resonance}

We first demonstrate REMPI by working with the pulse shape parameterization described above which produces a narrowband pulse whose central frequency can be scanned across the full laser spectrum.  

\begin{figure}[htbp]
    \centering	
    \figonecols{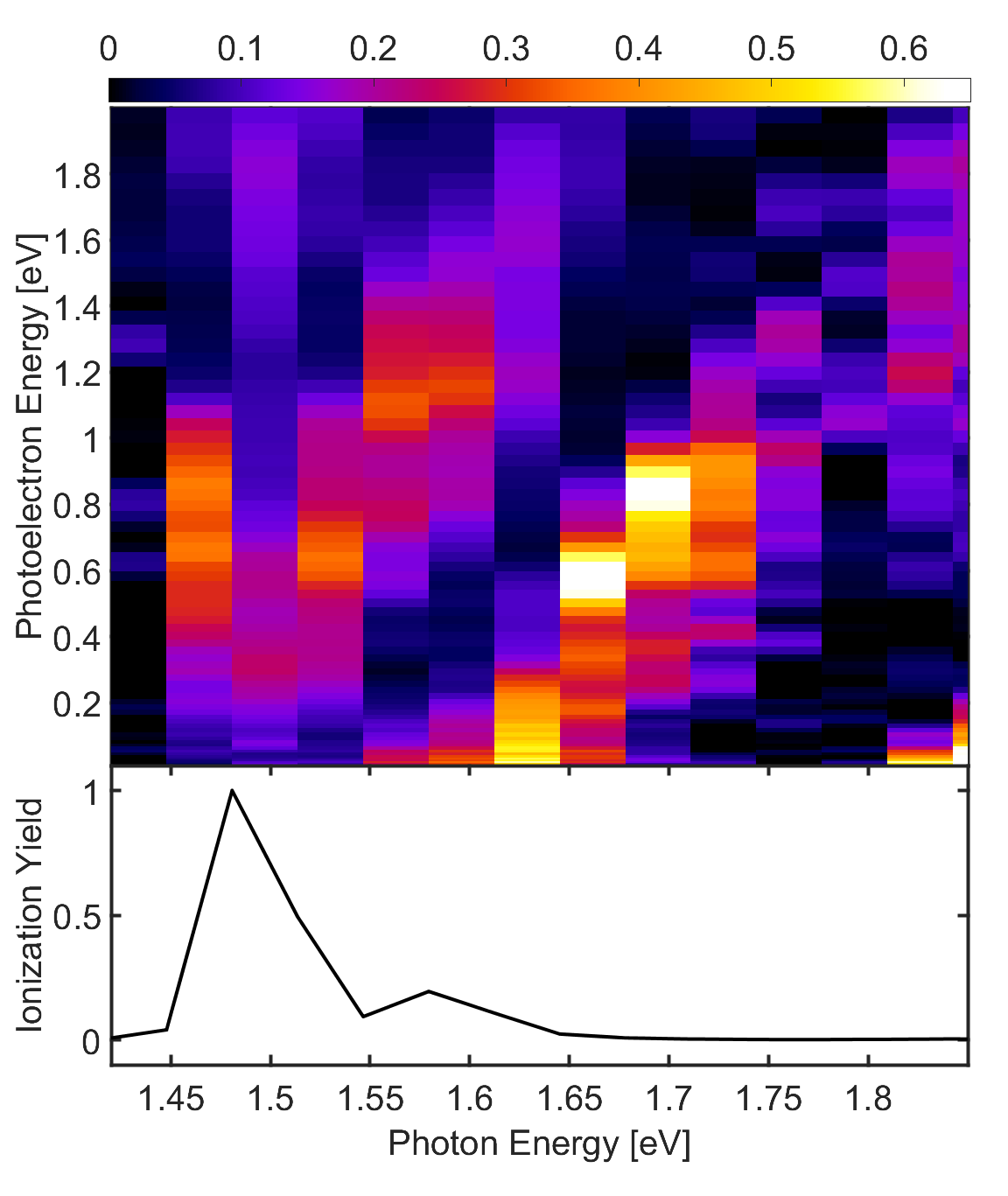}
    \caption{Normalized photoelectron yield as a function of photoelectron kinetic energy and photon energy (top panel) and photoelectron yield vs photon energy (bottom panel).}
    \label{fig:wavelength}
\end{figure}
Fig.~\ref{fig:wavelength} describes the experiment utilizing the $M_{WS}$ mask (Eq.~\ref{eq:maskWS}). Here $\Delta_\omega$ was fixed at $0.078$~rad/fs which corresponds to a $35$~fs pulse. The central optical frequency, $\omega_c$, was scanned across the optical spectrum and the PES at each frequency was measured. The top panel of Fig.~\ref{fig:wavelength} shows the normalized photoelectron yield as a function of photon energy and photoelectron energy. The bottom panel shows the un-normalized ionization yield (integral of the photoelectron spectrum) as a function of photon energy.  The yield clearly varies dramatically, with a peak near a photon energy of $1.5$~eV as a result of resonant enhancement of the muliphoton ionization.  This resonant enhancement also leads to an interruption in the diagonal line pattern of the top panel (near $1.50$~eV there is a noticeable smearing or doubling of the photoelectron peak), since near the resonant enhancement the photoelectron peak (energy, $K$) does not shift smoothly with photon energy, as one expects for the photoelectron energy associated with non-resonant ionization: 

\begin{equation}
    K = n\hbar\omega-I_p-U_p
\end{equation}
where $\hbar\omega$ represents the photon energy, n represents the multiphoton order, $I_p$ is the ionization potential (8.9~eV for thiophene), and $U_p$ is the ponderomotive potential ($< 0.5$~eV).

These measurements highlight the first, albeit trivial, condition for measuring coherence - there needs to be a strong multiphoton resonance coupling the ground state to the excited states of interest.

\subsection{Interference}
\begin{figure}[htbp]
    \centering	
    \figonecols{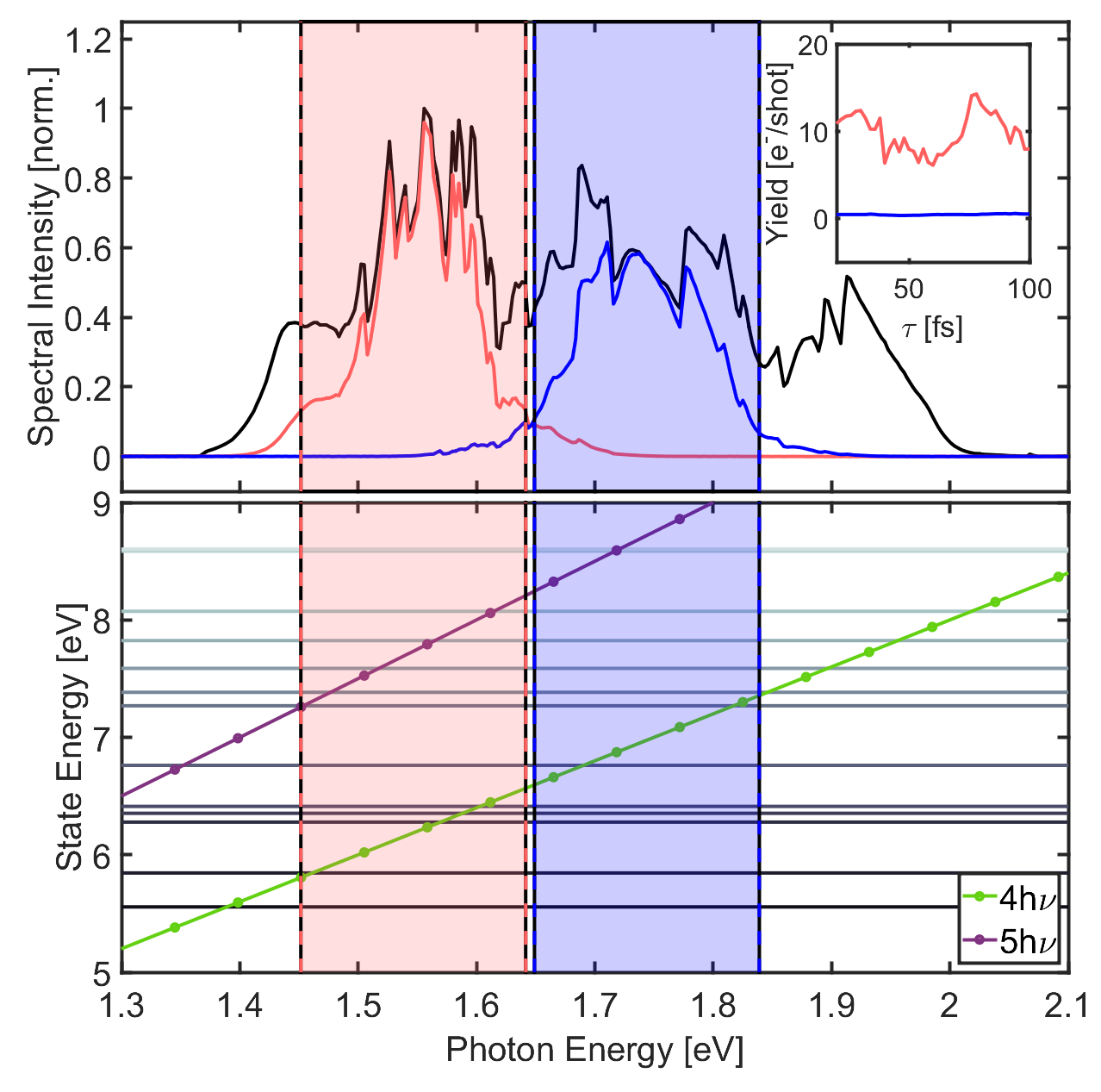}
    \caption{Top panel: optical spectra, with the full spectrum in black, and two narrowed spectra in red and blue, red ($1.55$~eV) and blue ($1.74$~eV) centered, respectively. Inset: pump-probe measurements using the two narrowed spectra color coded accordingly. Bottom panel: electronic structure calculated FC state energies (horizontal lines) compared with the 4 (green) and 5 (purple) photon excitation energies.}
    \label{fig:ppwavelength}
\end{figure}
Having established the importance of multiphoton resonances in enhancing the ionization yield, we now turn to measurements with the combined mask $M_{WPP}$ to perform a pump-probe measurement for narrowed spectra at different optical frequencies to establish the possibility of interference between two or more resonance enhancements. Here $\Delta_\omega$ was set to $0.18$~rad/fs which corresponds to $15$~fs pulses. The pump-probe measurement was performed for two central optical frequencies $2.35$~rad/fs ($1.55$~eV) and $2.65$~rad/fs ($1.74$~eV) . The top panel of Fig.~\ref{fig:ppwavelength} shows the optical spectrum as a function photon energy with the full spectrum in black, the spectrum centered at $1.55$~eV in red and $1.74$~eV in blue. The resulting photoelectron yield as a function of pump-probe delay is presented in the inset on the upper right. We see that the red centered spectrum resulted in a modulated yield while the blue centered spectrum is flat and nearly zero. This suggests that within the red spectrum are multiple resonant states which interfere with one another causing the modulation in the yield.  We note that while in principle the modulations could come from a vibrational coherence rather than an electronic one, further measurements described below for different locking frequencies make a clear case for electronic coherence.   

The bottom panel of Fig.~\ref{fig:ppwavelength} supports this idea by showing the Franck Condon (FC) energy for the excited states of thiophene between 5 and 9 eV determined by electronic structure calculations. The FC energies are independent of the applied photon energy so they correspond to flat lines which are shown purely for clarity so that the FC energy can be compared with the excitation energy for a 4 (green) and 5 (purple) photon process. The key point is that at no point within the blue spectral bandwidth (shaded in blue) do the 4 and 5 photon lines cross (come into resonance with) a state at the same photon energy. However, there are multiple photon energies ($1.45$~eV, $1.55$~eV, and $1.60$~eV for example) within the red spectral bandwidth (shaded in red), where both 4 and 5 photon resonance are possible for the same photon energy. This demonstrates both that there can be simultaneous $(n)^{\text{th}}$ and $(n+1)^{\text{th}}$ order resonances within the laser bandwidth, and if so, that these can lead to interference in the ionization yield. 

\section{Coherence}
\begin{figure}[htbp]
    \centering
    \figonecols{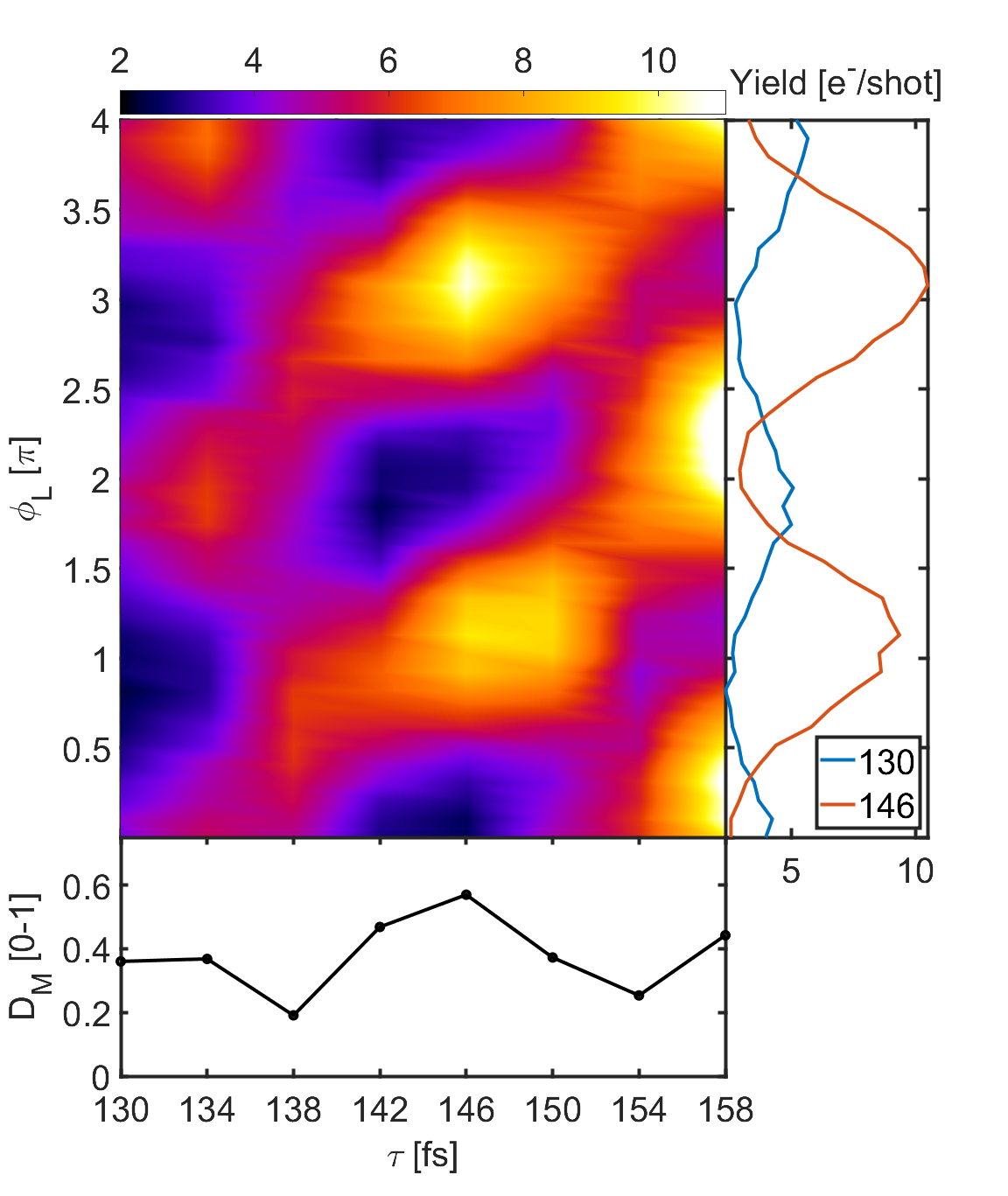}
    \caption{Measurements of the ionization yield of thiophene as a function of phase for different pump-probe delays.  Right and lower panels show the total yield  as a function of phase and the depth of modulation in the phase scan as a function of delay, respectively. The measurements highlight the phase dependence and the variation in depth of modulation with delay, illustrating the influence of the vibrational wave packet overlap on the measured coherence.}
    \label{fig:pump_probe}
\end{figure}

In order to explore the coherences in more detail, and confirm that they correspond to coherences between electronic states, we make use of the entire optical spectrum to perform a pump-probe measurement with the mask $M_{PP}$ (Eq.~\ref{eq:maskPP}). Rather than a traditional pump-probe where the delay $\tau$ is varied without independent control over the relative phase between pulses, we perform a delay locked phase scan where the delay is fixed and the relative pump-probe phase is scanned from $0-4\pi$. This measurement can be repeated for different delays to create the contour plot in Fig.~\ref{fig:pump_probe}, which shows the yield as a function of pump-probe delay and pump-probe phase.

A first obvious point from these measurements is that the yield clearly varies with phase for all of the measured delays, indicating that the molecular coherence in question is electronic.  The optical phase ($\phi_L$) dependence of the yield is consistent with Eq.~\ref{eq:final_yield_phase_tau}. While there is a clear phase dependence to the yield, one can see that the depth of modulation (D$_M$) varies with delay, highlighting the variation of vibrational overlap, which modulates the coherence, as described by Eq.~\ref{eq:coherence}. 

\begin{figure}[htbp]
    \centering	
    \figonecols{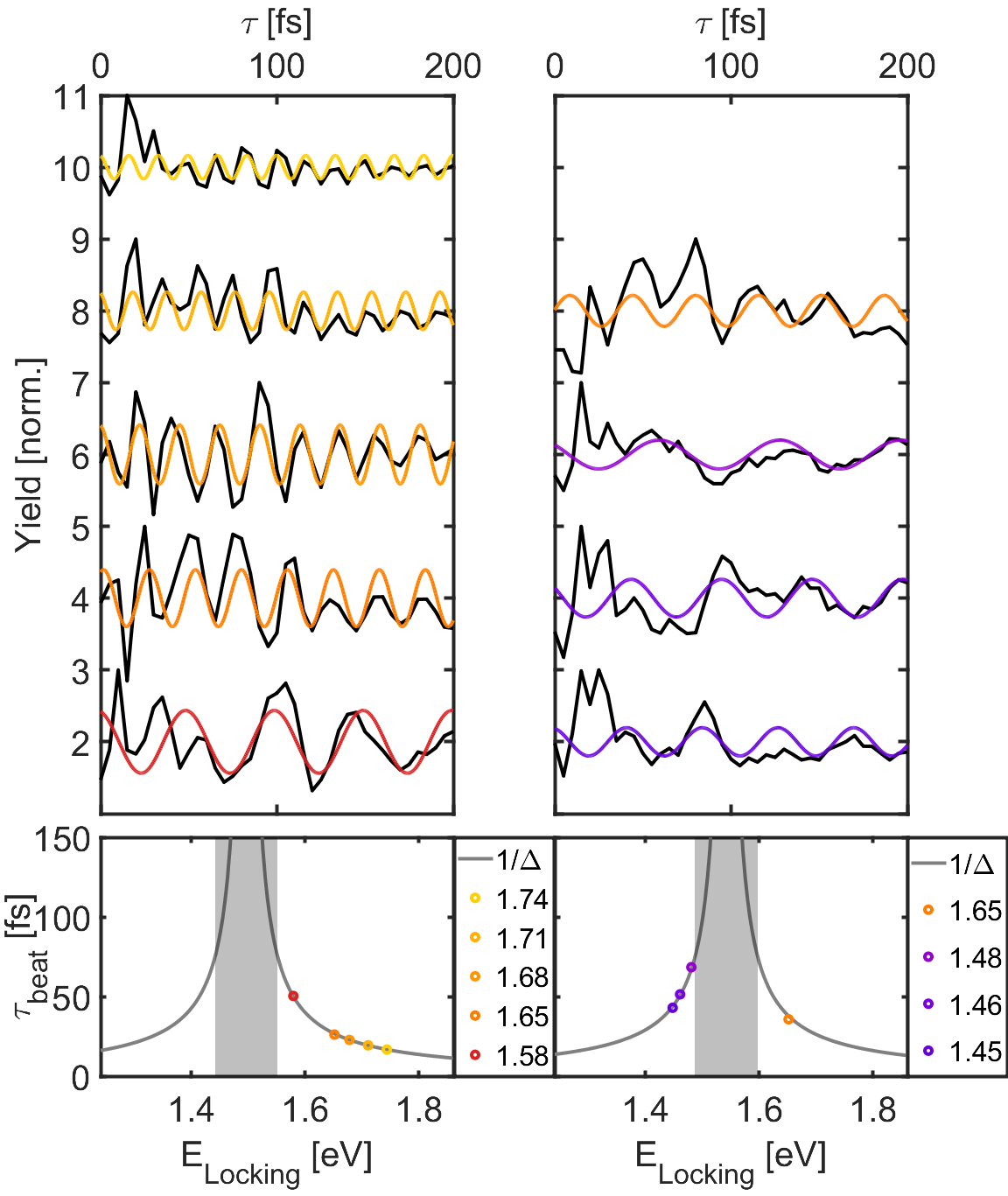}
    \caption{Pump-probe measurements for different locking frequencies. Top left panel: Pump-probe measurements for locking frequencies: $2.40$~rad/fs ($1.58$~eV), $2.51$~rad/fs ($1.65$~eV), $2.55$~rad/fs ($1.68$~eV), $2.60$~rad/fs ($1.71$~eV), and $2.65$~rad/fs ($1.74$~eV). Bottom left: Modulation period vs locking frequency for the measurements above. The solid line corresponds to Eq.~\ref{eq:beat} for a molecular resonance at $1.50$~eV.  Top right panel: Pump-probe measurements for locking frequencies: $2.20$~rad/fs ($1.45$~eV), $2.22$~rad/fs ($1.46$~eV), $2.25$~rad/fs ($1.48$~eV), and $2.51$~rad/fs ($1.65$~eV), measured on a different day with a slightly different peak intensity and laser spectrum. Bottom right: Modulation period vs locking frequency for the measurements above.The solid line corresponds to Eq.~\ref{eq:beat} for a molecular resonance at 1.55 eV.}
 \label{fig:locking_freq_scan}
\end{figure}

Fig.~\ref{fig:pump_probe} shows the yield as a function of phase and delay for a fixed locking frequency of $2.51$~rad/fs.  However, pump-probe measurements for different locking frequencies can help identify the energy difference between the electronic states leading to the modulations, as illustrated in Fig.~\ref{fig:locking_freq_scan}.  

Fig.~\ref{fig:locking_freq_scan} shows pump-probe measurements at different locking frequencies \cite{kaufman2023PRL}.  The left and right panels show pump-probe measurements conducted on different days with slightly different excitation conditions (similar but slightly different laser spectra and intensities) for several different locking frequencies (top panels).  The variation in modulation period ($\tau_{beat}$) with locking frequency highlights the electronic coherence, and is predicted by Eq.~\ref{eq:final_yield_phase_tau}:
\begin{equation}
    \tau_{beat} = \frac{2\pi}{(\omega_2-\omega_1)-\omega_{L}}    \label{eq:beat}
\end{equation}
where we have omitted the $\boldsymbol{R}$ dependence in $\omega_2$ and $\omega_1$, for the case of potentials which are roughly parallel.  The bottom panels in the figure show the measured modulation periods together with calculated curves for the case of resonances at $1.50$~eV and $1.55$~eV. The fact that the modulation periods in the yield for a given laser intensity and spectrum all lie along the curves in the bottom two panels indicates that coherences between pairs of states can dominate the interference in the ionization yield. However, the energy separation between the pair of states changes between measurements indicating that the interference is more complicated and can involve more than two states \footnote{While the subtleties of the the experimental conditions that lead to these differences are not well known these measurements are reproducible under similar conditions.}.  This is not surprising given the density of states at the 4 and 5 photon excitation level and the broad bandwidth of our laser pulses. So while only one pair of states needs to be resonant at the same photon energy to give rise to this coherence, it is possible to have multiple pairs of states resulting in the beating between coherences. For example, the measured resonance of $1.55$~eV in Fig.~\ref{fig:locking_freq_scan} is consistent with the expected resonances in the lower half of Fig.~\ref{fig:ppwavelength}. The $1.50$~eV is not as readily obvious. While it could be true that this pair of states is the main one excited due to the particular laser-molecular conditions, an alternative explanation is that the $1.50$~eV coherence comes from equal excitation of the $1.55$~eV and the $1.45$~eV pairs and arises as the beating of these pairs of pairs of states.


\section{Conclusion}
In conclusion, we have carried out and analyzed a phase-locked pump-probe measurements of electronic wave packets in thiophene which reveal long lived electronic coherences between excited states of the molecule. The measurements rely on simultaneous multiphoton resonances at different multiphoton orders (our analysis suggests that there are multiple pairs of electronic states contributing to the measurements), which are facilitated by using ultrabroadband laser pulses.  We illustrate how the coherences evolve with delay between the pump and probe pulses, highlighting the role of vibrations in modulating the coherence.   If one is able to measure the nuclei in coincidence with the photoelectrons (e.g. after double ionization with the probe pulse to produce the molecular dication, which can dissociate and produce pairs of fragment ions), then it would also be possible to further disentangle the electronic and nuclear dynamics, and acquire more insight into their coupled motion. 

\begin{acknowledgments}
This work was supported by National Science Foundation under award number 2110376. T. Rozgonyi acknowledges support from the National Research, Development and Innovation Fund of Hungary under grant numbers 2018-1.2.1-NKP-2018-00012 and SNN 135636.
\end{acknowledgments}

\section*{Conflict of Interest}
The authors declare that they have no conflicts of interest.

\section*{Data Availability}
The data that support the findings of this study are not publicly available at this time but may be obtained from the authors upon reasonable request.

\section*{Appendix}    
Here we provide some more details on the derivation of the expressions for the ionization yield as a function of pulse shape provided in the main text.  The laser field in time $E(t)$, can be defined as the product of an envelope $E_0(t)$ and a carrier with a frequency of $\omega_0$:
\begin{equation}
    E(t) = E_0(t)e^{-i\omega_0 t} +c.c.
    \label{Aeq:laser}
\end{equation}
To generate a double pulse with the pulse shaper we apply a mask function $M(\omega)$ to the optical spectrum of the form:
\begin{equation}
    M(\omega) = 1+A_Re^{i(\omega-\omega_L)\tau+i\phi_L}
    \end{equation}
This mask generates a pulse pair with controllable parameters: relative pulse amplitude $a$, locking frequency $\omega_L$, pump-probe delay $\tau$, and relative pump-probe phase $\phi_L$. The locking frequency sets the frequency at which there is always constructive interference in the optical spectrum.

Mathematically, the pulse shaper can be described by Fourier transforming the pulse $E(t)\rightarrow E(\omega)$, applying the mask $E'(\omega)=M(\omega)E(\omega)$, and Fourier transforming back $E'(\omega)\rightarrow E'(t)$. Where $E'(t)$ describes the desired shaped laser pulse in the time domain:
\begin{align}
    E'(t) &= \int_{-\infty}^{\infty} E'(\omega) e^{-i\omega t} d\omega \nonumber\\
    &= \int_{-\infty}^{\infty}dt E(\omega) e^{-i\omega t} + ae^{i(\phi_L-\omega_L\tau)}\int_{-\infty}^{\infty}dt E(\omega) e^{-i\omega (t-\tau)} \nonumber\\
    &= E(t) + A_RE(t-\tau)e^{i(\phi_L-\omega_L\tau)} \label{Aeq:laser_double}
\end{align}

Using Eq.~\ref{Aeq:laser}, we can define the pump and probe pulses by rewriting Eq.~\ref{Aeq:laser_double}. 
We define the pump pulse as $E_{\text{pu}}(t)=E_0(t)\cos(\omega_0 t)$ and the probe pulse as $E_{\text{pr}}(t)=E'_0(t-\tau)\cos(\omega_0 (t-\tau)-\phi)$, where we set $E'_0(t)=A_RE_0(t)$ and simplify the notation for the controllable parameters of the laser as $\phi = \phi_L-\omega_L\tau$.

We can now turn our attention to the experiment and how the laser interacts with the molecule. The pump pulse creates a coherent superposition of electronic states via multiphoton absorption. Reproduced from the main text Eq.~\ref{eq:coherence} describes the electronic coherence generated by the pump pulse in terms of the off diagonal elements of the density matrix: 
\begin{equation}
    \rho_{12}(\boldsymbol{R},t)=a_1 \chi_1(\boldsymbol{R},t) a_2^* \chi_2^*(\boldsymbol{R},t) e^{i\omega_{21}(\boldsymbol{R})t}\text{,}
    \label{Aeq:coherence}
\end{equation}
where $\omega_{21}(\boldsymbol{R}) = \omega_{2}(\boldsymbol{R})-\omega_1(\boldsymbol{R})$.

The ionization step can then be described by 
\begin{align} 
    Y(t)&=|a_1|^2|b_1|^2+|a_2|^2|b_2|^2 +b_1 b_2^*  \int d\boldsymbol{R} \rho_{12}(\boldsymbol{R},t) + c.c. \label{Aeq:yield_t}
\end{align}
where $b_1$ and $b_2$ represent $m$ and $m-1$ photon (with $m$=2 in this work) ionization amplitudes, which are proportional to the $m^{th}$ and ($m-1$)$^{th}$ powers of the probe pulse field respectively:
\begin{subequations}
\label{Aeq:bn}
\begin{align}
    b_1 &= Q_{1\text{f}}\Big(E'_0(t-\tau)e^{-i(\omega_0(t-\tau)-\phi)}+c.c.\Big)^m  \\
    b_2 &= Q_{2\text{f}}\Big(E'_0(t-\tau)e^{-i(\omega_0(t-\tau)-\phi)}+c.c.\Big)^{(m-1)}
\end{align}    
\end{subequations}
Here $Q_{1\text{f}}$ and $Q_{2\text{f}}$ represent field independent multi-photon matrix elements that can be written in terms of sums over off resonant intermediate states \cite{lunden2014model,kaufman2022numerical}. 

We consider contributions to the ionization yield that involve the minimum photon orders from states 1 and 2 to arrive at the same final state energy, and make the multiphoton rotating wave approximation. 
\comment{
 \begin{subequations}
\begin{align}
    b_1(t) &= Q_{1\text{f}}\big(RE_{0}(t-\tau)\big)^m\big(e^{-im(\omega_0(t-\tau)-\phi)}+c.c\big) \\
    b_2(t) &= Q_{2\text{f}}\big(RE_{0}(t-\tau)\big)^{(m-1)}\big(e^{-i(m-1)(\omega_0(t-\tau)-\phi)}+c.c\big) 
\end{align}    
\end{subequations}
}

Thus plugging Eqs.~\ref{Aeq:coherence} and \ref{Aeq:bn} into Eq.~\ref{Aeq:yield_t} and keeping only the relevant terms yields:
\begin{align}
    Y(t,\tau,\phi) &=|a_1|^2\Big|Q_{1\text{f}}(E'_0)^m(t-\tau)\Big|^{2} \nonumber \\
    &+|a_2|^2\Big|Q_{1\text{f}}(E'_0)^{(m-1)}(t-\tau)\Big|^{2} \nonumber\\
    &+a_1a_2^*Q_{2\text{f}}\big(E'_0(t-\tau)\big)^{m}e^{-im(\omega_0(t-\tau)-\phi)} \nonumber\\ &Q^*_{2\text{f}}\big(E'_0(t-\tau)\big)^{(m-1)}e^{i(m-1)(\omega_0(t-\tau)-\phi)} \nonumber \\
    &\int d\boldsymbol{R} \chi_1(\boldsymbol{R},t)\chi_2^*(\boldsymbol{R},t) e^{i\omega_{21}(\boldsymbol{R})t} + c.c. \label{Aeq:yield_t_tau_phi}
\end{align}
where, the final term can be simplified to:

\begin{align}
    D(t,\tau,\phi) &=a_1a_2^*Q_{1\text{f}}Q^*_{2\text{f}} \big(E'_0(t-\tau)\big)^{(2m-1)} 
    e^{-i(\omega_0(t-\tau)-\phi)} \nonumber \\ 
    &\int d\boldsymbol{R} \chi_1(\boldsymbol{R},t)\chi_2^*(\boldsymbol{R},t) e^{i\omega_{21}(\boldsymbol{R})t} + c.c. \label{Aeq:D_t_tau_phi}
\end{align}

The experiment measures the time integrated yield, $Y(\tau) = \int_{t_1}^{t_2} Y(t) dt$ where $t_1$ is the time after the pump pulse has turned off and $t_2$ is the time after the probe pulse has turned off. Our interest is in the electronic coherence so we can focus on this third term, Eq.~\ref{Aeq:D_t_tau_phi} which carries all of the phase information. Performing the integral over time results in:
\begin{align}
    D(\tau,\phi) &= a_1a_2^*Q_{1\text{f}}Q^*_{2\text{f}} e^{i(\omega_0\tau+\phi)} \int_{t_1}^{t_2} dt \big(E'_0(t-\tau)\big)^{(2m-1)} \nonumber\\
    &\int d\boldsymbol{R} \chi_1(\boldsymbol{R},t)\chi_2^*(\boldsymbol{R},t) e^{i(\omega_{21}(\boldsymbol{R})-\omega_0)t}  + c.c.\label{Aeq:D_tau}
\end{align}

Upon rearranging the previous equation to account for the integral over time we make note of two phase terms that arise: $\omega_0\tau+\phi$ and $(\omega_{21}(\boldsymbol{R})-\omega_0)t$. The first of these is time independent and arises from the probe pulse. It describes the phase advance of the laser as a function of pump-probe delay plus the controllable phase $\phi$. The second phase comes from the initial excitation by the pump pulse and describes the phase advance of the coherence with respect to that of the laser.


The pulse duration of the laser pulses are short with respect to the molecular dynamics, and given that the ionization signal is a another factor of $\sqrt{2m-1}$ than the laser pulse we continue the analysis in the impulsive limit where $(E'_0(t-\tau))^{(2m-1)}=(E'_0)^{(2m-1)}\delta(t-\tau)$ \footnote{While the strict impulsive limit, where the field approaches a delta function, is in tension with the multiphoton rotating wave approximation, we note that our experiments operate in a regime where the pulse is short compared to the nuclear dynamics in question, but long compared to the detuning period of the counter rotating terms in the RWA: 1/$(\omega_{21}+n\omega_0)$}\cite{kaufman_2020_AdiabaticEliminationStrongfield}. Thus, $t\rightarrow \tau$ and we can rewrite the total phase as $\omega_0\tau+\phi+\omega_{21}(\boldsymbol{R})\tau-\omega_0\tau$ or $\omega_{21}(\boldsymbol{R})\tau+\phi$. This simplifies Eq.~\ref{Aeq:D_tau} to:
\begin{align}
    D(\tau,\phi) &= a_1a_2^*Q_{1\text{f}}Q^*_{2\text{f}} (E'_0)^{(2m-1)}\int d\boldsymbol{R} \chi_1(\boldsymbol{R},\tau)\chi_2^*(\boldsymbol{R},\tau) \nonumber\\
    &e^{i(\omega_{21}(\boldsymbol{R})\tau+\phi)} + c.c.
\end{align}

At last we can write the measured yield as:
\begin{align}
    Y(\tau,\phi) &= |a_1|^2|Q_{1f}(E'_0)^m|^2+|a_2|^2|Q_{2f}(E'_0)^{m-1}|^2 \nonumber\\
    &+a_1a_2^*Q_{1\text{f}}Q^*_{2\text{f}}(E'_0)^{(2m-1)} \int d\boldsymbol{R} \chi_1(\boldsymbol{R},\tau)\chi_2^*(\boldsymbol{R},\tau) \nonumber
    \\
    &e^{i((\omega_{21}(\boldsymbol{R})-\omega_L)\tau+\phi_L)} + c.c.\label{Aeq:final_yield}
\end{align}
which maintains sensitivity to the three decoherence mechanisms described in Eq. \ref{Aeq:coherence}. However, the dephasing term is now modified by the applied laser phase, which we have reintroduced as the two types of controllable laser parameters. This gives rise to the two types of pump-probe measurements that can be performed: (1) delay scan with fixed phase, and (2) phase scan with fixed delay.


\comment{\begin{align}
    D(\tau) &= Q_{1\text{f}} Q^*_{2\text{f}}a_1a_2^*e^{-i\phi} \int_{t_1}^{t_2} dt \big(E_0(t-\tau)\big)^{(2m-1)}e^{i\omega_0(t-\tau)} \nonumber \\
    &\int d\boldsymbol{R} \chi_1(\boldsymbol{R},t)\chi_2^*(\boldsymbol{R},t) e^{i\omega_{21}(\boldsymbol{R})t} + c.c. \nonumber
\end{align}

replacing $t'=t-\tau$ yields

\begin{align}
    D(\tau) &= Q_{1\text{f}} Q^*_{2\text{f}}a_1a_2^* e^{-i\phi}\int_{t_1}^{t_2} dt' \big(E_0(t')\big)^{(2m-1)}e^{i\omega_0t'} \nonumber \\
    &\int d\boldsymbol{R} \chi_1(\boldsymbol{R},t'+\tau)\chi_2^*(\boldsymbol{R},t'+\tau) e^{i\omega_{21}(t'+\tau)} + c.c. \nonumber \\
    &= Q_{1\text{f}} Q^*_{2\text{f}}a_1a_2^*e^{i\phi}e^{i\omega_{21}\tau} \int_{t_1}^{t_2} dt' \big(E_0(t')\big)^{(2m-1)}e^{i(\omega_{21}+\omega_0)t'} \nonumber \\
    &\int d\boldsymbol{R} \chi_1(\boldsymbol{R},t'+\tau)\chi_2^*(\boldsymbol{R},t'+\tau) + c.c. \nonumber
\end{align}

This leads to a measured time and phase dependent yield which can be written as:
\begin{align}
      Y_{pp}(\tau,\phi)&=|a_1|^2|Q_{1f}E_0^m|^2+|a_2|^2|Q_{2f}E_0^{m-1}|^2 \label{eq:dication_yield2} 
      \\ 
      &+ Q_{1\text{f}} Q^*_{2\text{f}} E_0^{(2m-1)}e^{i\phi}  \int d\boldsymbol{R} \rho_{12}(\boldsymbol{R},\tau) + c.c. \nonumber
  \end{align}


\begin{table}[]
\centering
\caption{These three values can be shown to have double peaked structure in the wavelength scans. The $1.50$eV ($363$THz) corresponds to the resonance calculated based on the electronic coherence in the submitted manuscript. The $1.55$eV corresponds to the resonance calculated by the electronic coherence in the middle panel of Fig. \ref{fig:Thio_Repeat_Comp}. }
\label{tab:my-table}
\begin{tabular}{|c|c|c|c|}
\hline
h$\nu$                                                    & eV                                                            & eV                                                            & eV                                                            \\ \hline
1                                                         & \begin{tabular}[c]{@{}c@{}}1.45\\ (350THz=857nm)\end{tabular} & \begin{tabular}[c]{@{}c@{}}1.50\\ (363THz=826nm)\end{tabular} & \begin{tabular}[c]{@{}c@{}}1.55\\ (375THz=800nm)\end{tabular} \\ \hline
4                                                         & 5.8                                                           & 6                                                             & 6.2                                                           \\ \hline
5                                                         & 7.25                                                          & 7.5                                                           & 7.75                                                          \\ \hline
\begin{tabular}[c]{@{}c@{}}6\\ 6h$\nu$-8.9eV\end{tabular} & \begin{tabular}[c]{@{}c@{}}8.7\\ -0.2\end{tabular}            & \begin{tabular}[c]{@{}c@{}}9\\ 0.1\end{tabular}               & \begin{tabular}[c]{@{}c@{}}9.3\\ 0.4\end{tabular}             \\ \hline
\begin{tabular}[c]{@{}c@{}}7\\ 7h$\nu$-8.9eV\end{tabular} & \begin{tabular}[c]{@{}c@{}}10.15\\ 1.25\end{tabular}          & \begin{tabular}[c]{@{}c@{}}10.5\\ 1.6\end{tabular}            & \begin{tabular}[c]{@{}c@{}}10.85\\ 1.95\end{tabular}          \\ \hline
\end{tabular}
\end{table}}

\bibliographystyle{unsrt}
\bibliography{references}

\end{document}